\documentclass[12pt]{article}

\usepackage[preprint, nonatbib]{neurips_2023}

\usepackage[utf8]{inputenc} 
\usepackage[T1]{fontenc}    
\usepackage{hyperref}       
\usepackage{url}            
\usepackage{booktabs}       
\usepackage{amsfonts}       
\usepackage{nicefrac}       
\usepackage{microtype}      
\usepackage{xcolor}         

\usepackage{fancyhdr}

\usepackage{algorithm}
\usepackage{algpseudocode}

\hypersetup{colorlinks, linkcolor=blue, citecolor=blue, urlcolor=blue}

\usepackage{amsfonts,dsfont}
\usepackage{hyperref}
\usepackage{color}
\usepackage{amsmath, amssymb}
\usepackage{amsthm}
\usepackage{bm}
\usepackage{stmaryrd}
\usepackage{color}
\usepackage{comment}
\usepackage{geometry}
\allowdisplaybreaks

\usepackage{graphicx}
\usepackage{caption}
\usepackage{subcaption}

\usepackage{tabularx}

\usepackage{caption}

\newcommand{\todo}[1][\null]{\ensuremath{\clubsuit}}

\newcommand{\noprint}[1]{}

\theoremstyle{definition}
\theoremstyle{definition}
\theoremstyle{definition}
\theoremstyle{definition}
\theoremstyle{definition}
\theoremstyle{definition}
\theoremstyle{definition}
\theoremstyle{definition}
\theoremstyle{definition}

\title{Physics-informed neural networks\\ for tsunami inundation modeling}

\author{%
  R\"udiger Brecht \\
  Department of Mathematics\\
  University of Hamburg\\
  Hamburg, Germany \\
  \texttt{ruediger.brecht@uni-hamburg.de} \\
  \And
  Elsa Cardoso-Bihlo \\
  Department of Mathematics and Statistics\\
  Memorial University of Newfoundland\\
  St. John’s, NL, A1C 5S7, Canada \\
  \texttt{ecardosobihl@mun.ca} \\
  \And
  Alex Bihlo \\
  Department of Mathematics and Statistics\\
  Memorial University of Newfoundland\\
  St. John’s, NL, A1C 5S7, Canada \\
  \texttt{abihlo@mun.ca} \\
}

\begin{document}

\maketitle

\vspace{0.25cm}

\begin{abstract}
We use physics-informed neural networks for solving the shallow-water equations for tsunami modeling. Physics-informed neural networks are an optimization based approach for solving differential equations that is completely meshless. This substantially simplifies the modeling of the inundation process of tsunamis. While physics-informed neural networks require retraining for each particular new initial condition of the shallow-water equations, we also introduce the use of deep operator networks that can be trained to learn the solution operator instead of a particular solution only and thus provides substantial speed-ups, also compared to classical numerical approaches for tsunami models.
We show with several classical benchmarks that our method can model both tsunami propagation and the inundation process exceptionally well.
\end{abstract}

\section{Introduction}
Tsunamis are a single or a series of giant waves that are generated by displacement of water masses and can be triggered by different mechanisms: (underwater) landslides, underwater earthquakes, and earth surface movements adjacent to the ocean~\cite{robke2017tsunami,voit1987tsunamis}. The modelling of tsunamis pose significant challenges. Generating landslide tsunamis presents several difficulties related to the collection of enough data, insufficiently understood interactions between solid and fluid components, and uncertainty in longer duration landslide processes which impacts wave generation and propagation. A further factor in modeling these type of tsunamis is the dependence on several physical parameters including seafloor topography and landslide variables \cite{dias2014modelling}. Tsunamis triggered by underwater earthquakes are directly linked to questions of displacement of the earth floor and energy transfer to the water column. Finally, tsunamis caused by earth surface movements near the ocean involves rare events like volcano eruptions or meteor strikes, with smaller spatial scales in comparison to the other types of tsunamis~\cite{di2011forecasting,paris2015source}. That is, tsunamis encompass several length scales from planetary to small-scale turbulence, presenting significant challenges in the simulation of their dynamics and impacts on coastal areas.

Many equations have been used for tsunami modeling, including the shallow-water equations, the Boussinesq equations, the Serre--Green--Naghdi equations and the Navier--Stokes equations~\cite{marr20a}. Recent years have also seen a shift away from numerical approaches solving the governing equations of free surface flow to data-driven approaches based on machine learning~\cite{liu21a}. 

It is also possible to combine data-driven and numerical approaches, and this is the realm of \textit{scientific machine learning}. In this paper we demonstrate that tsunami inundation modeling based on the shallow-water equations is feasible using physics-informed machine learning strategies. Specifically we consider the case of classical \textit{physics-informed neural networks} learning particular solutions of the shallow-water equations, and the case of \textit{physics-informed DeepONets}, which learn the solution operator of a class of initial--boundary value problems for the shallow-water equations. The proposed method is entirely meshless, does not rely on numerical discretization to approximate the derivatives in the shallow-water equations, and thus also does not require any specific inundation model, making it potentially much easier to implement than traditional numerical approaches. We demonstrate with several classical benchmarks that both physics-informed neural networks and DeepONets can accurately model the moving wet--dry interface in the shallow-water equations as required for tsunami inundation modeling.

The further organization of this paper is the following. Section~\ref{sec:RelatedWork} reviews the pertinent literature on tsunami modeling using numerical and machine learning approaches. We present a concise introduction to physics-informed neural networks and physics-informed DeepONets in Section~\ref{sec:PINNs}. The framework for using physics-informed neural networks for the shallow-water equations with variable bottom topography is discussed in Section~\ref{sec:PINNsSWE}. Section~\ref{sec:NumericalResults} presents the numerical results obtained on several classical benchmarks of relevance for tsunami inundation modeling. The final Section~\ref{sec:Conclusions} contains a summary of the results of this paper together with some discussion about potential next research avenues within this area of scientific machine learning.

\section{Related work}\label{sec:RelatedWork}
  
Tsunami modeling has been an active field of research for several decades, owing to the devastating effects of large-scale tsunamis such as the 2004 Indian ocean or the 2011 Tohoku tsunami, with several more pre-historical tsunamis still detectable in geological features and sediments, see e.g.~\cite{papa14a}. Advancements in numerical methods and increasing computational resources and data storage have helped propel the numerical study of several aspects of tsunamis with high fidelity as a more practical option to in-lab physical experimentation. 
The first tsunami simulations were initiated in Japan \cite{aida1969numerical} and proved to be an effective approach in modeling the initial formation of tsunamis caused by several types of disturbances \cite{lopez2015advanced,ulrich2019coupled}, the travel of the tsunami wave and its interaction with the varying depths and topography of the ocean~\cite{bone18a,titov2005real}, and the end stage of the tsunami by flooding coastal areas \cite{lunghino2020protective,prasetyo2019physical}. The tsunami-shore interaction has been of particular interest using one-dimensional and two-dimensional idealized models to study the role of isolated components such as underwater topography and vegetation in this context \cite{lunghino2020protective,okal2016sequencing}. 

For studying tsunami propagation and inundation a variety of governing equations of wave dynamics and discretization methodologies have been considered. These include finite-difference methods~\cite{goto97a,lyne02a,tito97Ay,tito98a}, finite elements and discontinuous Galerkin methods~\cite{bokh05a,bone18a,lai12a,vate15a}, meshless methods~\cite{bihl17a,brec18a} and smoothed particle hydrodynamics based approaches~\cite{stge14a}. Most of these approaches considered use either the shallow-water equations, Boussinesq-type equations or the Navier--Stokes equations. A summary of the some of the models that have been proposed over the past few decades is described in~\cite{marr20a}.

As with most other fields of science, also tsunami modeling has seen an increase in machine learning and data-driven approaches. Rather than modelling the physical processes underlying tsunami generation, propagation and inundation directly, data-driven approaches aim to model relationships of interest. These relationships include, for example, the connection between open ocean water displacement and the runup at a fixed location or the relationship between low- and high-resolution simulations, using machine learning methodologies. Such supervised approaches can yield impressive results in often a fraction of the time it takes to run a full physical tsunami simulation model~\cite{maki21a,muli20a,muli22a}. For a review on data-driven approaches with different machine learning techniques, see~\cite{liu21a}. 

There is also a middle ground between purely numerical approaches and machine learning techniques relying solely on data, which is referred to as \textit{scientific machine learning}. Here one aims to embed an inductive bias, often given in the form of differential equations or fundamental physical quantities such as symmetries and conservation laws, into a machine learning-based algorithm. For a comprehensive overview of scientific machine learning, see the survey by \cite{cuom22a}. A sub-field of scientific machine learning dedicated to solving differential equations using neural networks, known as \textit{physics-informed neural networks}, has gained considerable interest. These approaches date back to the seminal work of~\cite{laga98a} and were popularized more recently in~\cite{rais19a}. Physics-informed neural networks are attractive as they can be used to solve differential equations directly, similar to other numerical approaches such as finite differences, finite elements, and finite volume methods. They approximate derivatives using automatic differentiation~\cite{bayd18a} rather than numerical differencing, and as such are truly meshless methods. Additionally, physics-informed neural networks can also be used to solve inverse problems, such as identifying differential equations that underlie some observational data~\cite{rais19a}.

While physics-informed neural networks (PINNs) have garnered significant attention in recent years as a promising alternative to traditional numerical methods, earlier approaches~\cite{laga98a,rais19a} face some limitations. These limitations include difficulties in obtaining solutions for long time intervals~\cite{kris21a,wang22b} and their rather long training times, which can make them challenging to apply to real-time forecasting problems  ~\cite{brec23b}. Owing to the popularity of physics-informed neural networks, however, several mitigation strategies were proposed, which include multi-model approaches~\cite{bihl22a,kris21a}, improved training and optimization strategies~\cite{bihl23a, mccl22a, wang22a}, preservation of geometric properties~\cite{aror23a,card23a} and operator learning techniques~\cite{lu21a,wang23a}.

Especially the operator learning approach, also referred to as \textit{physics-informed DeepONets} is a notable generalization to physics-informed neural networks. Rather than training a physics-informed neural network for each particular solution of a system of differential equations, the procedure aims at directly learning the solution operator instead. This means that once the network has learned the solution operator, it can be used to produce the solution for arbitrary initial conditions, which considerably speeds up the time in which a solution can be computed. Data-driven DeepONets are behind some of the most impressive state-of-the-art results in numerical meteorology today~\cite{bi23a,lam23a}. The improvement in computational times make DeepONets especially suitable for time-critical tasks such as tsunami forecasting.

Related applications to what will be studied in this paper, physics-informed neural networks have already been used to solve the shallow-water equations on the sphere~\cite{bihl22a}, and they have found applications in numerical meteorology as well~\cite{brec23b}. There are also several applications to hydrodynamical problems involving the shallow-water equations, see~\cite{anel21a,huan23a,leit21a,qi24a}. Specifically, the shallow-water equations for some classical hydrodynamical problems, such as the dam break problem in one and two dimensions were considered in~\cite{anel21a}, and a one-dimensional test case with fully submersed variable bottom topography but without inundation was considered in~\cite{leit21a}. The authors of~\cite{huan23a} studied the shallow-water equations in the Lagrangian framework with applications to flat bottom topographies, and~\cite{qi24a} considered applications to river flooding.

\section{A primer on physics-informed neural networks}\label{sec:PINNs}

Physics-informed neural networks were introduced in~\cite{laga98a} and later popularized in~\cite{rais19a}. We present a short overview of the method here, and refer to the papers~\cite{bihl22a, cuom22a, laga98a, lu21a, rais19a} for a more extended presentation. Given a system of differential equations over the spatio-temporal domain $\Omega=[0,t_{\rm f}]\times \Omega_{\rm s}$ where $\Omega_s\subset \mathbb{R}^d$,
\begin{align}\label{eq:InitialBoundaryValueProblem}
\begin{split}
    &\Delta^l(t,\mathbf{x},\mathbf{u}_{(n)}) = 0,\quad l=1,\dots, L,\quad t\in[0,t_{\rm f}],\mathbf{x}\in\Omega,\\
    &\mathsf{I}^{l_{\rm i}}(\mathbf{x},\mathbf{u}_{(n_{\rm i})}|_{t=0})=0,\quad l_{\rm i}=1,\dots,L_{\rm i},\quad \mathbf{x}\in\Omega,\\
    &\mathsf{B}^{l_{\rm b}}(t,\mathbf{x},\mathbf{u}_{(n_{\rm b})}) = 0,\quad l_{\rm b} = 1,\dots, L_{\rm b}, \quad t\in[0,t_{\rm f}], \mathbf{x}\in\partial \Omega,
\end{split}
\end{align}
where $t$ is the time variable, $\mathbf{x}=(x_1,\dots, x_d)$ denotes the tuple of spatial independent variables, $\mathbf{u}=(u^1,\dots, u^q)$ are the dependent variables and the subscript $u_{(n)}$ refers to the tuple of all derivatives of dependent variables with respect to the independent variable of order not exceeding $n$. The initial and boundary conditions are encoded in the initial value and boundary value operators, $\mathsf{I}=(\mathsf{I}^1,\dots,\mathsf{I}^{L_{\rm i}})$ and $\mathsf{B}=(\mathsf{B}^1,\dots,\mathsf{B}^{L_{\rm b}})$, respectively. For the system of shallow-water equations, a system of evolution equation, using Dirichlet boundary conditions, we have $\mathsf{I}=\mathbf{u}|_{t=0}-\mathbf{u}_0$ and $\mathsf{B}=\mathbf{u}-\mathbf{g}(t,\mathbf{x})$, for fixed vector functions $\mathbf{u}_0=(u_0^1(\mathbf{x}),\dots,u_0^q(\mathbf{x}))$ and $\mathbf{g}(t,\mathbf{x})=(g^1(\mathbf{x}),\dots,g^q(\mathbf{x}))$, respectively.

The idea behind physics-informed neural networks is to learn a global solution interpolant parameterized using a deep neural network. Usually, a fully connected multi-layer perceptron is used here. Let $\boldsymbol{\theta}$ denote the parameters (weights and biases) of a neural network $\mathcal{N}^{\boldsymbol{\theta}}$, then the goal of a physics-informed neural network is to learn the parameterization $\mathbf{u}^{\boldsymbol{\theta}}=\mathcal{N}^{\boldsymbol{\theta}}(t,\mathbf{x})$, where ideally $\mathbf{u}^{\boldsymbol{\theta}}(t,x) \approx \mathbf{u}(t,x)$. To learn the neural network weights~$\boldsymbol{\theta}$, a finite collection of \textit{collocation points} is sampled over the spatio-temporal domain for which the solution should be learned. 

Specifically, denote by $\{(t_\Delta^i,\mathbf{x}_\Delta^i)\}_{i=1}^{N_\Delta}$ the \textit{differential equations collocation points}, by $\{(t_{\rm i}^i,\mathbf{x}_{\rm i}^i)\}_{i=1}^{N_{\rm i}}$ the \textit{initial data collocation points} and by $\{(t_{\rm b}^i,\mathbf{x}_{\rm b}^i)\}_{i=1}^{N_{\rm b}}$ the \textit{boundary data collocation points}. Then, the neural network is trained to minimize the composite loss function
\begin{subequations}\label{eq:PINNloss}
\begin{equation}\label{eq:PINNlossComposite}
\mathcal L(\boldsymbol{\theta}) = \mathcal L_{\Delta}(\boldsymbol{\theta}) + \gamma_{\rm i}\mathcal L_{\rm i}(\boldsymbol{\theta}) + \gamma_{\rm b}\mathcal L_{\rm b}(\boldsymbol{\theta}),
\end{equation}
where the physics (or differential equation) loss, the initial value loss and the boundary value loss are given, respectively, by
\begin{align}\label{eq:PINNlossComponents}
\begin{split}
    &\mathcal L_{\Delta}(\boldsymbol{\theta}) = \frac{1}{N_{\rm \Delta}}\sum_{i=1}^{N_\Delta}\sum_{l=1}^L|\Delta^l(t_{\rm \Delta}^i,\mathbf{x}_{\rm \Delta}^i,\mathbf{u}_{(n)}^{\boldsymbol{\theta}}(t_{\rm \Delta}^i,\mathbf{x}_{\rm \Delta}^i))|^2,\\
    &\mathcal L_{\rm i}(\boldsymbol{\theta}) = \frac{1}{N_{\rm i}}\sum_{i=1}^{N_{\rm i}}\sum_{l_{\rm i}=1}^{L_{\rm i}}|\mathsf{I}^{l_i}(t_{\rm i}^i,\mathbf{x}_{\rm i}^i,\mathbf{u}^{\boldsymbol{\theta}}_{(n_{\rm i})}(t_{\rm i}^i, \mathbf{x}_{\rm i}^i))|^2,\\
    &\mathcal L_{\rm b}(\boldsymbol{\theta}) = \frac{1}{N_{\rm b}}\sum_{i=1}^{N_{\rm b}}\sum_{l_{\rm b}=1}^{L_{\rm b}}|\mathsf{B}^{l_{\rm b}}(t_{\rm b}^i,\mathbf{x}_{\rm b}^i,\mathbf{u}_{(n_{\rm b})}^{\boldsymbol{\theta}}(t_{\rm b}^i,\mathbf{x}_{\rm b}^i))|^2,
\end{split}
\end{align}
\end{subequations}
and $\gamma_{\rm i}$ and $\gamma_{\rm b}$ are positive constants (loss weight parameters). Upon minimization of the physics-informed loss function $\mathcal L(\boldsymbol{\theta})$ the weights of the neural network will be tuned such that ideally $\mathbf{u}^{\boldsymbol{\theta}}(t,x)$ will satisfy the initial--boundary value problem at the sampled collocation points.

A key attribute of physics-informed neural networks is that the required derivatives $\mathbf{u}^{\boldsymbol{\theta}}_{(n)}$, $\mathbf{u}^{\boldsymbol{\theta}}_{(n_{\rm i})}$ and $\mathbf{u}^{\boldsymbol{\theta}}_{(n_{\rm b})}$ in the differential equation, initial condition and boundary condition are computed using \textit{automatic differentiation}~\cite{bayd18a}. This is possible since the neural network can be differentiated with respect to its input parameters $t$ and $\mathbf{x}$ provided that differentiable activation functions, such as hyperbolic tangents or sigmoid activations, are used throughout the network. Crucially, automatic differentiation is readily available in modern deep learning toolkits such as \texttt{JAX}, \texttt{PyTorch} and \texttt{TensorFlow} as automatic differentiation is being used to compute the derivatives of the loss function with respect to the neural network parameters, which is required for the gradient-based optimization approaches that are used to train neural networks~\cite{lecu15a}. This way of computing derivatives makes physics-informed neural networks completely meshless, as no discretization grids are required to compute derivatives. As a consequence, the method is well-suited for solving differential equations on complicated domains, such as the sphere~\cite{bihl22a} or arbitrary embedded manifolds~\cite{tang22a}.

It was first noted in~\cite{laga98a} that the initial and boundary conditions can alternatively be enforced as a \textit{hard constraint} by using a specific solution ansatz for $\mathbf{u}^{\boldsymbol{\theta}}$. Notably, using the ansatz
\[
\mathbf{u}^{\boldsymbol{\theta}}(t,\mathbf{x}) = F_{\rm i}(t,\mathbf{x})\mathbf{u}_0(\mathbf{x}) + F_{\rm b}(t,\mathbf{x}) + F_{\rm nn}(t,\mathbf{x})\mathcal N^{\boldsymbol{\theta}}(t,\mathbf{x}),
\]
the functions $F_{\rm i}$, $F_{\rm b}$ and $F_{\rm nn}$ can be chosen such that $\mathbf{u}^{\boldsymbol{\theta}}(t_0,\mathbf{x})=\mathbf{u}_0$ for all $\mathbf{x}\in\Omega$ and $\mathbf{u}^{\boldsymbol{\theta}}(t,\mathbf{x})=\mathbf{g}(t,\mathbf{x})$ for all $(t,\mathbf{x})\in[0,t_{\rm f}]\times\partial \Omega$, i.e.\ that both the initial and spatial Dirichlet boundary conditions are automatically satisfied. In this case, the training of a physics-informed neural network reduces to minimizing the differential equation loss. For further information, see~\cite{brec23a}. The advantage of this hard constrained approach over the \textit{soft-constrained} approach presented in~\eqref{eq:PINNloss} is that in the latter there is no guarantee that the initial and boundary conditions are exactly satisfied. As we will see below, however, for solutions that are not differentiable everywhere, using the hard-constrained approach can slow down learning, as the hard-constrained neural network will have to learn to offset any moving wet--dry interfaces.

While physics-informed neural networks have been the subject of considerable study recently, they face two main challenges: \textit{i)} They take a lot of time to train; and \textit{ii)} they cannot be used for long time integrations, as they tend to converge to trivial solutions in these cases. Both of these shortcomings can be overcome by learning the solution operator of a differential equation, rather than the solution itself. That is, let $\mathfrak{G}(\mathbf{u}_0)$ be the solution operator of system~\eqref{eq:InitialBoundaryValueProblem}, such that $\mathbf{u}(t,\mathbf{x}) = \mathfrak{G}(\mathbf{u}_0(\mathbf{x}))$ is the solution of~\eqref{eq:InitialBoundaryValueProblem} with initial condition $\mathbf{u}_0$ for all $t\in[0,t_{\rm f}]$ and all $\mathbf{x}\in\Omega$.

Practically, physics-informed deep operator networks are trained as physics-informed neural networks with an extra input vector obtained by sampling the initial condition $\mathbf{u}_0=\mathbf{u}_0(\mathbf{x})$ at a finite number $N_{\rm s}$ of fixed \textit{sensor points}, i.e.\ $\{\mathbf{u}_0(\mathbf{x}_s)\}_{s=1}^{N_{\rm s}}$ is passed into the network along with the independent variables $t$ and $\mathbf{x}$. The most classical way of combining the sampled initial condition and the independent variables is through a two sub-neural network architecture with the first sub-neural network being referred to as the \textit{branch net}, processing the sampled initial condition, and the second sub-neural network being referred to as the \textit{trunk net}, processing the independent variables. The branch and trunk networks produce each a $k$-dimensional output vector, $\mathbf{b}=(b_1,\dots, b_k)$ and $\mathbf{t}=(t_1,\dots,t_k)$, with $k$ being a hyper-parameter, which are then scalar-multiplied to obtain the following operator representation approximation
\[
\mathfrak{G}(\mathbf{u}_0(\mathbf{x})(t,\mathbf{x}) \approx \mathbf{b}(\mathbf{u}_0(\mathbf{x}_1),\dots,\mathbf{u}_0(\mathbf{x}_{N_{\rm s}})\cdot \mathbf{t}(t, \mathbf{x}).
\]
This representation is inspired by the theoretical results derived in~\cite{chen95a}, who showed that any continuous nonlinear operator between two Banach spaces can be represented in this manner, where $\mathbf{b}$ and $\mathbf{t}$ are the outputs of a single layer perceptron. This result parallels the result by Cybenko~\cite{cybe89a} showing that neural networks with a single hidden layer are universal function approximators. For more details, see~\cite{chen95a, lu21a, wang23a}. In practice, more than a single hidden layer is being used, giving rise to deep operator networks. 

Generalizations to the classical branch and trunk network representation of deep operator networks have been investigated extensively in the past few years, see, e.g.~\cite{lee23a}. In the following we will use a version of NOMAD~\cite{seid22a}, for our operator network architecture, which passes the sampled initial condition at the sensor points through a suitable branch network, the output of which is then concatenated with the independent variables (the inputs to the trunk network), after which this concatenated vector is passed through a standard multi-layer perceptron, then outputting the dependent variables.

\section{Physics-informed neural networks for the shallow-water equations with bottom topography}\label{sec:PINNsSWE}

The nonlinear shallow-water equations are a set of partial differential equations describing the propagation of a depth-averaged fluid under the assumption that the horizontal length scale is much larger than the vertical length scale. They can be written as a generalized conservation law of the form
\begin{equation}\label{eq:ShallowWaterEquations}
\boldsymbol{\rho}_t + \mathbf{F}_x + \mathbf{G}_y = \mathbf{S},
\end{equation}
where $\boldsymbol{\rho}=(h,hu, hv)^{\rm T}$ is the transport vector of total mass and horizontal momentum, $\mathbf{F}=(hu, hu^2 + \tfrac{1}{2}gh^2, huv)^{\rm T}$ and $\mathbf{G}=(hv, huv, hv^2 +\tfrac{1}{2}gh^2)^{\rm T}$ are the flux vectors, and $\mathbf{S}=(0,-ghb_x, -ghb_y)^{\rm T}$ is the source term. The height of the water column of constant density over the time-independent bottom topography $b=b(x,y)$ is denoted by $h=h(t,x,y)$, and the two-dimensional, depth-averaged fluid velocity vector is given by $(u,v)^{\rm T}=(u(t,x,y),v(t,x,y))^{\rm T}$. The gravitational constant is $g$. Here and in the following we use the usual subscript notation to denote derivatives with respect to the independent variables $t$, $x$ and $y$. Additional source terms can be added to account for bottom friction or diffusion processes, but will not be considered in the present study. For more details, see e.g.~\cite{anel21a, kais11a}. We consider initial--boundary value problems for the shallow-water equations~\eqref{eq:ShallowWaterEquations} on the spatio-temporal domain $[0,t_{\rm f}]\times[L^x_{\rm min},L^x_{\rm max}]\times[L^y_{\rm min},L^y_{\rm max}]$ using Dirichlet boundary conditions.  In the following, we will also consider the one-dimensional version of the shallow-water equations~\eqref{eq:ShallowWaterEquations} in which case $h=h(t,x)$, $u=u(t,x)$, $v=0$ and $b=b(x)$.

We aim to solve the shallow-water equations~\eqref{eq:ShallowWaterEquations} using physics-informed neural networks. That is, we design a neural network interpolant of the form $\mathbf{z}^{\boldsymbol{\theta}}=(u^{\boldsymbol{\theta}},v^{\boldsymbol{\theta}},h^{\boldsymbol{\theta}})^{\rm T} = \mathcal N^{\boldsymbol{\theta}}(t,x,y)$, such that~\eqref{eq:ShallowWaterEquations} are satisfied along with any appropriate initial and boundary conditions.

A main complication when numerically solving the shallow-water equations for problems with a moving wet--dry interface is the time-dependent inundation and drying algorithm that is typically required. This has given rise to a host of numerical methods on various discretization stencils, see e.g.~\cite{bihl17a,bokh05a,bone18a,brec18a,goto97a,lai12a,lyne02a,tito97Ay,tito98a,vate15a}, all of which implement specific wetting and drying algorithms. Crucially, no such approach is required when using physics-informed neural networks, as the required derivatives here are not computed using numerical, grid based discretization, but using automatic differentiation. This alleviates the need for using a specific moving boundary, or wetting--drying approach. Instead, we just distribute the required collocation points throughout the entire spatio-temporal computational domain $[0,t_{\rm f}]\times\Omega$, and train a global solution interpolant for the shallow-water equations. As such, wetting and drying happens automatically without the need for any specific additional inundation algorithm. We believe that this fact, together with the meshless nature of physics-informed neural networks, enabling finding the solution for complicated spatial domains straightforward, makes this approach exceptionally well-suited for tsunami inundation modeling.

In practice, we found it advantageous to have the neural network learn the velocity components $u$ and $v$ rather than the momentum variables $uh$, and $vh$ since having to obtain the former from the latter by dividing with $h$ was prone to numerical instability for small $h$. This is in part due to neural networks typically being trained using single precision arithmetic. It is also well-known that deep neural networks suffer from training difficulties when the input parameters vary greatly~\cite{glor10a}. As such, we apply a static normalization layer as a first layer in the neural network which normalizes the input variables $t$, $x$ and $y$ so that each of them to lie in the interval $[0,1]$. To account for the different magnitudes in the output variables $h$, $u$ and $v$, we non-dimensionalize the differential equation, initial value and boundary value losses in~\eqref{eq:PINNlossComponents} with the characteristic time scales and magnitudes of the dependent variables. This ensures that each loss term in~\eqref{eq:PINNlossComponents} contributes at equal rate to the composite physics-informed neural network loss~\eqref{eq:PINNlossComposite}.

\section{Numerical results}\label{sec:NumericalResults}

In the following we present several classical benchmarks for the shallow-water equations with application to inundation simulation. Specifically, Section~\ref{subSec:PINNs} is devoted to the training of physics-informed neural networks for the shallow-water equations, while Section~\ref{subSec:DeepONets} considers the case of directly learning the solution operators for particular cases of interest in shallow-water modeling. For both types of problems we found experimentally that for all results where there is a wet--dry interface present in the initial condition it is advantageous to enforce the initial condition as a soft constraint on the physics-informed loss function rather than as a hard constraint, which is in line with the observations made in~\cite{anel21a}. This is due to the initial condition being non-differentiable at the wet--dry interface. As this interface is moving throughout the simulation, the neural network has to learn to offset the non-differentiable initial condition which slows down learning. This is due to the spectral bias in neural networks~\cite{raha19a} which learn low-frequency solution components faster than high-frequency components, and as such struggle with learning the moving wet--dry interface.

We use \texttt{TensorFlow} 2.16 to train our physics-informed neural networks, using a single NVIDIA RTX 8000 GPU. We use the Adam optimizer~\cite{king14a} with a learning rate of $10^{-3}$ using mini-batch gradient descent with 10 batches per epoch and train all neural networks until the loss saturates. The codes for reproducing the results can be found on \texttt{GitHub} upon publication of the article.\footnote{\url{https://github.com/abihlo/PINNTsunami}}

\subsection{Physics-informed neural networks}\label{subSec:PINNs}

We consider several classical benchmarks for the shallow-water equations with variable bottom topography using physics-informed neural networks. That is, we train neural networks to act as surrogate models for the solution to the shallow-water equations in the form of global  interpolants, defined for the entire spatio-temporal domain of interest over which the numerical solutions are sought. 

All examples below use a four hidden layer fully connected neural network with 40 units per layer employing a hyperbolic tangent activation function. Hyperparameter tuning has revealed that a change in this baseline architecture did not substantially change the results obtained, indicating that the main source of error of the trained network is due to the optimization method being used; further improvements can be expected upon using more sophisticated optimization approaches, such as meta-learned optimization~\cite{bihl23a} or quasi-Newton methods such as L-BFGS~\cite{cuom22a}, but we refrain from using these methods here for the sake of simplicity of the exposition.

\subsubsection{Lake-at-rest solution}

The lake-at-rest solution to the shallow-water equations describes the scenario where the fluid velocity is zero and the total water height remains constant, corresponding to the steady-state solution $(u,v,h+b) = (0,0,\mathrm{const})$. Numerical schemes for the shallow-water equations that can maintain this steady-state solution exactly are referred to as \textit{well-balanced} in the tsunami literature, cf.~\cite{audu04a,brec18a,vate15a,xia13a}.

Achieving a well-balanced solution to the shallow-water equations using physics-informed neural networks is conceptually different as the solution is modelled by a global solution interpolator that is learned using an optimization strategy. As such, it is readily possible to inject the well-balanced property into the ansatz for this interpolant, so that the neural network surrogate solution for the shallow-water equations satisfies the late-at-rest solution even before training the neural network. This is useful, as the lake-at-rest solution is rarely of interest in itself but is considered to make sure that a numerical method does not introduce spurious modes in the discretization procedure. Well-balancedness of the global physics-informed neural network interpolant solution can be achieved using both soft- and hard-constrained initial conditions. We consider the one-dimensional case as the two-dimensional case is completely analogous. 

For hard-constrained initial conditions, we can use the ansatz
\begin{align*}
&h^{\boldsymbol{\theta}}(t,x) = h_0(x) + \frac{t}{t_{\rm f}}\frac{(x-L_{\rm min})(x-L_{\rm max})}{L_{\rm max}-L_{\rm min}}\mathcal{N}_h^{\boldsymbol{\theta}}(t,x),\\ &u^{\boldsymbol{\theta}}(t,x) = u_0(x) + \frac{t}{t_{\rm f}}\frac{(x-L_{\rm min})(x-L_{\rm max})}{L_{\rm max}-L_{\rm min}}\mathcal{N}_u^{\boldsymbol{\theta}}(t,x).
\end{align*}
If the weights of the neural network $\mathcal{N}^{\boldsymbol{\theta}}=(\mathcal{N}^{\boldsymbol{\theta}}_u,\mathcal{N}^{\boldsymbol{\theta}}_h)$ are initialized arbitrarily small then the solution will remain arbitrarily close to being well-balanced while satisfying the initial and Dirichlet boundary conditions.

\begin{figure}[!ht]
\centering
\begin{subfigure}[b]{\textwidth}
\includegraphics[width=0.45\textwidth]{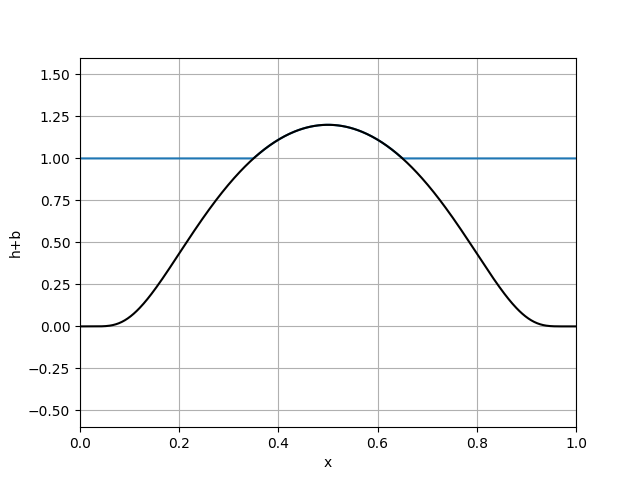}
\includegraphics[width=0.45\textwidth]{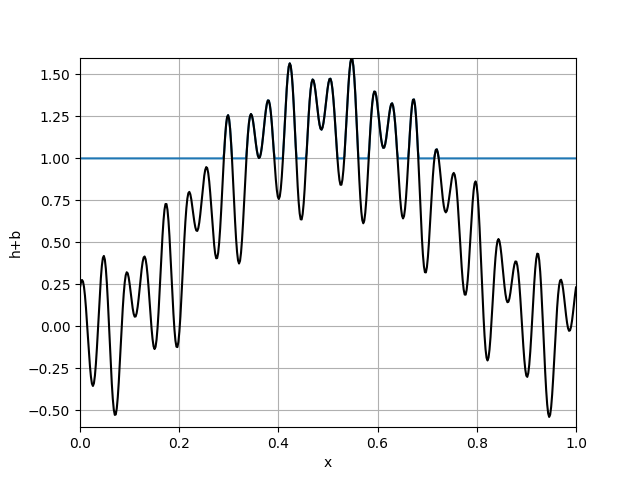}
\end{subfigure}
\caption{Lake at rest solution for a smooth and a noisy bottom topography using a well-balanced physics-informed neural network.}
\label{fig:LakeAtRest}
\end{figure}

For the soft-constraint approach we note that a neural network $\mathcal{N}^{\boldsymbol{\theta}}(t,x)$ is a multi-layer perceptron of the form
\[
\mathcal N^{\boldsymbol{\theta}}(t,x) = \sigma(W_n\sigma(W_{n-1}\cdots \sigma(W_1(t,x)^{\rm T}+\mathbf{b}_1) + \mathbf{b}_2)\cdots + \mathbf{b}_n).
\]
As such, to have the initial approximation $h^{\boldsymbol{\theta}}(t,x) = \mathcal N^{\boldsymbol{\theta}}_h(t,x)$ satisfy $h^{\boldsymbol{\theta}}(t,x) = h_0(x)$ we have to set all the weight matrices $W_i^{h}$, $i\in\{1,\dots,n\}$ and all bias vectors $\mathbf{b}^{h}_i$, $i\in\{1,\dots,n-1\}$ to arbitrary small numbers, and set the bias $\mathbf{b}^{h}_n=(b^{h}_n)$ as a function of the form $b^{h}_n(x)=h_0(x)-b(x)$. For the neural network $\mathcal N^{\boldsymbol{\theta}}_u$ we also set $b^{u}_n$ as an arbitrary small number. We show the corresponding initialized, well-balanced solutions for a smooth and a noisy bottom topography in Fig.~\ref{fig:LakeAtRest}.

Both ansatzes will guarantee that the neural network solution \textit{before} training is well-balanced. However, initializing the weights of a neural network to arbitrarily small numbers slows down learning and as such is not desirable for non-stationary solutions. We have found in practice, however, that the above initialization is still valuable for reducing the initial value loss component of the physics-informed neural network loss~\eqref{eq:PINNloss}, by initializing the network capable of capturing a steady state solution, which is typical for the onset of tsunami inundation simulations. 

Still, we note here that physics-informed neural networks approximate the solution to a differential equation in a fundamentally different manner to standard numerical methods. A global space-time neural network interpolator is learned to satisfy the shallow-water equations at all spatio-temporal collocation points, rather than an iterative time-stepper as is being used for standard numerical methods. As such, we believe that the notion of well-balancedness in physics-informed neural networks is less pertinent than for traditional numerical methods, for which well-balancedness is required to avoid spurious modes. 

\subsubsection{One-dimensional oscillatory flow in a parabolic basin}

Thacker~\cite{thac81a} introduced a series of exact solutions for the shallow-water equations with parabolic bottom topography. In the one-dimensional case it was found that for the bottom topography of the form $b=h_0(x/a)^2$ a particular solution of the form
\[
h_{\rm a} = h_0 -\frac{B^2}{4g}(1+\cos2\omega t) - \frac{Bx}{2a}\sqrt{\frac{8h_0}{g}}\cos(\omega t),\quad u_{\rm a} = \frac{Ba\omega}{\sqrt{2h_0g}}\sin\omega t,
\]
can be found, where $\omega=\sqrt{2gh_0/a}$. We follow the numerical setup from~\cite{brec18a,vate15a} and choose $h_0=10$, $a=3000$ and $B=5$ and solve the shallow-water equations on the spatial domain $[-5000,5000]$ for one complete period $t_{\rm f}=2\pi/\omega$. 

To train the associated physics-informed neural network, we use 10,000 collocation points for the differential equation, and 100 collocation points for the initial condition. No boundary collocation points had to be used for this specific case, as the numerical solution at the boundary remains constant, which is well-captured by our trained neural network.

\begin{figure}[!ht]
\centering
\begin{subfigure}[b]{\textwidth}
\includegraphics[width=0.33\textwidth]{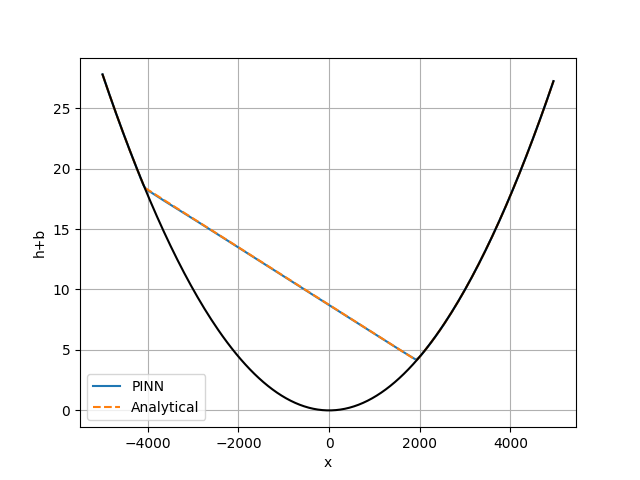}
\includegraphics[width=0.33\textwidth]{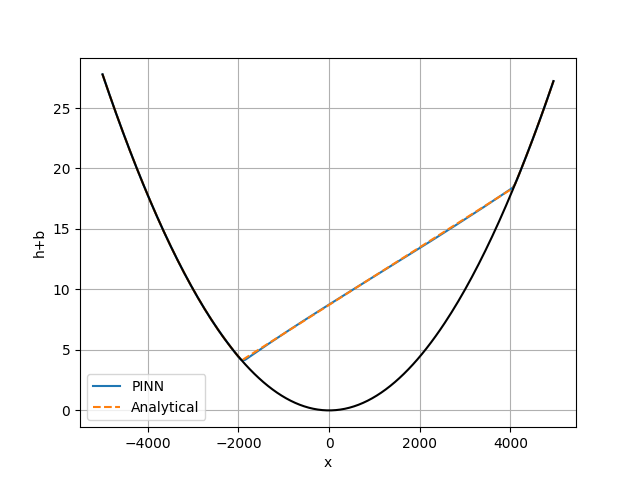}
\includegraphics[width=0.33\textwidth]{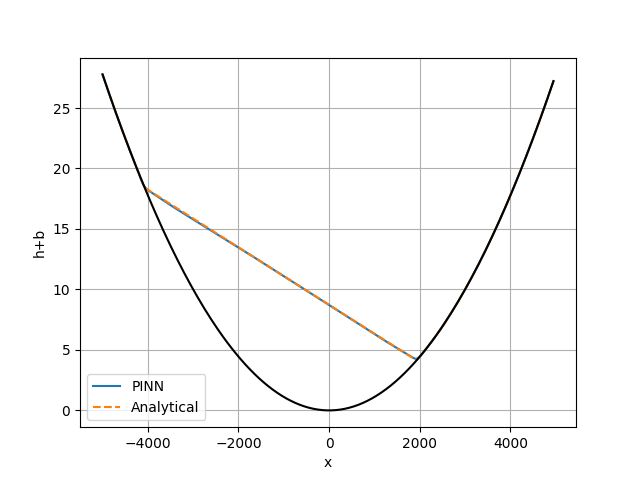}
\end{subfigure}
\caption{Oscillatory flow in a one-dimensional parabolic basis using a physics-informed neural network. The period of the solution is $2\pi/\omega$ and the results are shown at $t=0$ \textit{(left)}, $t=\pi/\omega$ \textit{(middle)} and $t=2\pi/\omega$ \textit{(right)}.}
\label{fig:ParabolicBowl1D}
\end{figure}

The numerical solution found using physics-informed neural networks is presented in Figure~\ref{fig:ParabolicBowl1D}. The numerical results obtained verify the success of the physics-informed neural network in capturing the moving wet--dry interface, with the numerical solution closely aligning with the exact solution.

\subsubsection{Carrier--Greenspan solution}

Carrier and Greenspan~\cite{carr58a} derived a class of exact solutions for the one-dimensional shallow-water equations for non-breaking waves inundating a sloping beach. They used the fact that for a linear bottom topography $b=x$, the one-dimensional shallow-water equations are linearizable using a hodograph transformation. Specifically, they introduced the coordinate transformation 
\begin{align*}
&t = \frac{\lambda}{2} - u,\quad x = \phi_\lambda -\frac{\sigma^2}{16}-\frac{u^2}{2}\\
&u = \frac{\phi_\sigma}{\sigma},\quad h = \frac{\phi_\lambda}{4} - \frac{u^2}{2} 
\end{align*}
so that the shallow-water equations can be reduced to the single second-order linear partial differential equation
\[
(\sigma \phi_\sigma)\sigma -\sigma\phi_{\lambda\lambda} = 0.
\]
They then proceeded to construct the class of particular solutions of the form
\begin{equation}\label{eq:CarrierGreenspan}
\phi(\lambda,\sigma)=AJ_0(\sigma)\cos(\lambda),
\end{equation}
where $J_0$ denotes the Bessel function of the first kind. This solution can be used to reconstruct a solution for the shallow-water equations in terms of the original variables as long as the Jacobian $\partial(t,x)/\partial)(\lambda,\sigma)$ does not vanish, which holds for for $\sigma>0$ (with $\sigma=0$ denoting the moving shoreline in the new coordinate system) and $A\leqslant 1$, see also~\cite{bokh05a}. In the Carrier and Greenspan solution, $g=1$.

We train physics-informed neural networks for the cases of $A=0.5$, $A=1.0$, which were presented in~\cite{carr58a}. The initial condition is the wave at the deepest point and the integration is carried out until the time of maximal run-up $t=\pi/2$. We use the exact solution as boundary condition at the open ocean boundary on the left side of the spatial domain.

\begin{figure}[!ht]
\centering
\begin{subfigure}[b]{\textwidth}
\includegraphics[width=0.49\textwidth]{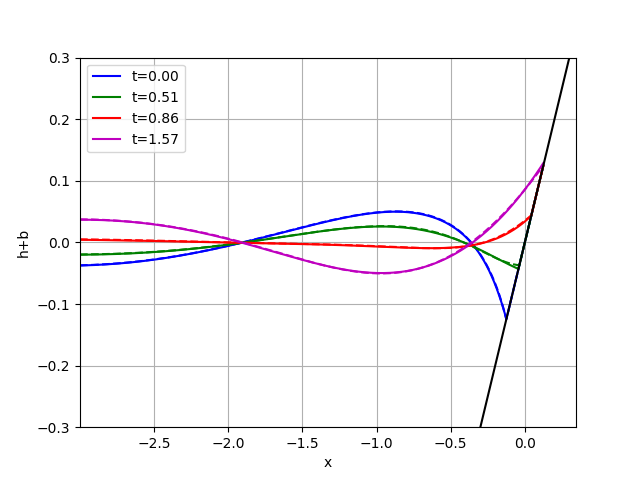}
\includegraphics[width=0.49\textwidth]{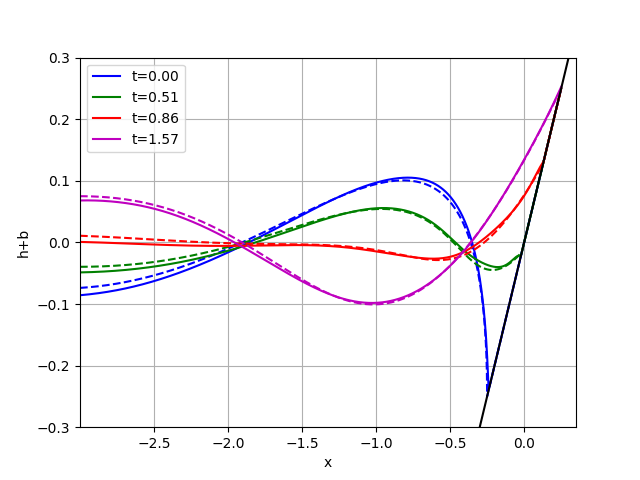}
\end{subfigure}
\caption{Carrier--Greenspan solution~\eqref{eq:CarrierGreenspan} for $A=0.5$ \textit{(left)} and $A=1.0$ \textit{(right)}. Solid lines denote the physics-informed neural network solution, dashed lines the analytical solution.}
\label{fig:CarrierGreenspan}
\end{figure}

Figure~\ref{fig:CarrierGreenspan} shows the results of the physics-informed neural network solutions, along with the errors compared to the exact solution. A total of 20,000 differential equations collocation points, and 200 initial and 200 boundary collocation points were used. In both cases was the physics-informed neural network able to accurately predict the maximal run-up of the incoming wave. 

\subsubsection{Two-dimensional oscillatory flow in a parabolic basin}

\begin{figure}[!ht]
\centering
\begin{subfigure}[b]{\textwidth}
\includegraphics[width=0.33\textwidth]{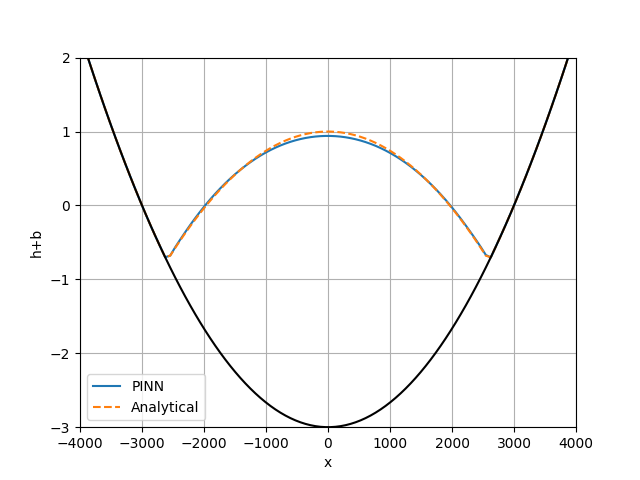}
\includegraphics[width=0.33\textwidth]{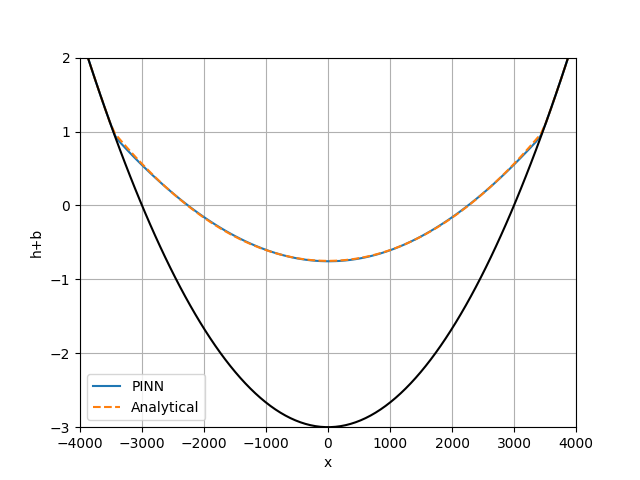}
\includegraphics[width=0.33\textwidth]{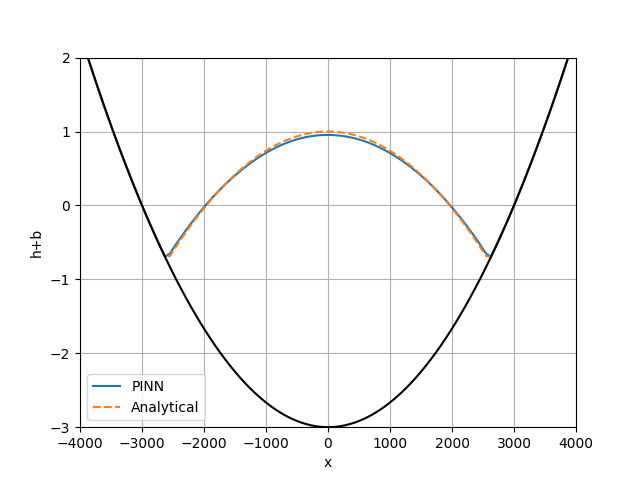}
\end{subfigure}
\caption{Oscillatory flow in a parabolic basis using a physics-informed neural network. The period of the solution is $2\pi/\omega$ and the results are shown at $t=0$ \textit{(left)}, $t=\pi/\omega$ \textit{(middle)} and $t=2\pi/\omega$ \textit{(right)} at the cross section $y=0$.}
\label{fig:ParabolicBowl2D}
\end{figure}

Thacker~\cite{thac81a} considered the case of the two-dimensional shallow-water equations in a parabolic bowl with bottom topography $b=b_0(r^2/L^2-1)$, with $r=\sqrt{x^2+y^2}$. A particular solution for this bottom topography is given by
\begin{align*}
    &h_{\rm a}(t,x,y) = b_0\left(\frac{\sqrt{1-A^2}}{1-A\cos\omega t} - 1 - \frac{r^2}{L^2}\left(\frac{1-A^2}{(1-A\cos\omega t)^2}\right)\right)\\
    &(u_{\rm a},v_{\rm a}) = \frac{A\omega\sin\omega t}{2-2A\cos\omega t}(x,y),
\end{align*}
where $\omega = \sqrt{8gb_0}/L$ and $A=((b_0+z_0)^2-b_0^2)/((b_0+z_0)+b_0^2)$. The numerical setup for this benchmark is similar to~\cite{lai12a}, in that we use $b_0=3$, $z_0=1$, $L=3000$. We solve this problem for one complete period $t_{\rm f}=2\pi/\omega$, and present the numerical result in Figure~\ref{fig:ParabolicBowl2D}.

As in the related one-dimensional benchmark, also for this two-dimensional example is the numerical solution obtained from using physics-informed neural networks highly accurate and able to capture the correct period of the solution.

\subsection{Physics-informed DeepONets}\label{subSec:DeepONets}

Having trained several physics-informed neural networks in the previous Section~\ref{subSec:PINNs}, we now turn our attention to the operator learning case. Here, neural networks have to be optimized to learn the solution operator of the shallow-water equations. This implies that a class of suitable initial conditions has to be chosen, for which the solution operator is being learned. For the purpose of this paper we consider a narrow class of initial conditions for each of the following examples, which is obtained by using the exact solution at various times $t$ as the initial condition. However, since this exact solution is only evaluated at finitely many sensor points $N_{\rm s}$, also numerical solutions obtained from classical numerical solvers such as those discussed in Section~\ref{sec:RelatedWork} could be used as initial conditions to train the physics-informed DeepONets.

The one-dimensional physics-informed deep operator network in this section use a simplified NOMAD architecture,~\cite{lee23a,seid22a}, with the inputs to the branch network being concatenated to the inputs of the trunk network after which the concatendated vector is passed through a standard multi-layer perceptron with 8 hidden layers and 40 units per layer. The two-dimensional physics-informed DeepONet uses the NOMAD architecture, with the branch network using a convolutional neural network to process the two-dimensional gridded input functions for $u_0$, $v_0$ and $h_0$. A total of four convolutional layers with 8, 16, 32 and 64 filters of size $3\times3$ and stride 2 is used for each input dependent variable to the network, the results of which are globally averaged and pooled, and then concatenated with the inputs to the trunk network. The resulting concatenated vector is then processed by the standard trunk net of the form of a multi-layer perceptron with four layers with 40 units each. We use a convolutional neural network for the branch net in the two-dimensional case rather than a fully connected network, as the two-dimensional gridded sensor points would lead to a rather wide multi-layer perceptron. The hyperbolic tangent activation function is being used throughout the network. The one-dimensional DeepONet uses $N_{\rm s}=100$ sensor points, the two-dimensional DeepONet uses $N_{\rm s}= 32\times 32$ sensor points.

\subsubsection{Dam break problem}\label{subsubSec:DamBreakProblem}

Since the classical dam break problem has been studied in multiple papers using physics-informed neural networks, see e.g.~\cite{anel21a,huan23a,leit21a}, here we consider this problem using physics-informed deep operator networks. Specifically, we consider the classical one-dimensional dam break problem for the shallow-water equations for the case of a flat bottom topography $b=0$. 

The analytical solution for this benchmark is given by 
\begin{align*}
    &gh_{\rm a}(t,x) = \left\{\begin{array}{ll}
     a_0^2 & \textup{if}\quad x-x_0 < -a_0t\\
     \frac{1}{9}\left(2a_0-(x-x_0)/t\right)^2 & \textup{if}\quad -a_0t < x-x_0 < 2a_0 t \\
     0 & \textup{if}\quad x-x_0 > 2a_0t
    \end{array}\right.\\
    &u_{\rm a}(t,x) = \left\{\begin{array}{ll}
     0 & \textup{if}\quad x-x_0 < -a_0t\\
     \frac{2}{3}\left(a_0+(x-x_0)/t\right)^2 & \textup{if}\quad -a_0t < x-x_0 < 2a_0 t \\
     0 & \textup{if}\quad x-x_0 > 2a_0t
    \end{array}\right.,
\end{align*}
where $a_0=\sqrt{gh_0}$, see~\cite{bokh05a} for further details. 

For training the physics-informed deep operator network for this problem we randomly sample a total of 10,000 initial conditions from the class of problems where $h_0$ is drawn from a uniform random distribution over the interval $[0.1,1]$, and $x_0$ from a uniform random distribution over the interval $[-\pi/4, \pi/4]$. The exact solution for various values of $t\in[0,2]$ is used as an initial condition for training the DeepONets, which aim to learn the solution operator over the spatio-temporal domain $[-\pi,\pi]\times[0,\Delta t]$ for a time-step of $\Delta t = 1$. We set $g=1$ for this benchmark.

\begin{figure}[!ht]
    \centering
\begin{subfigure}[b]{\textwidth}
\includegraphics[width=0.49\textwidth]{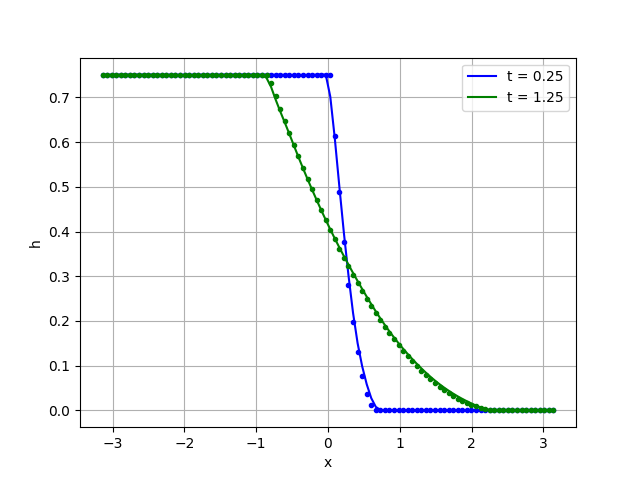}
\includegraphics[width=0.49\textwidth]{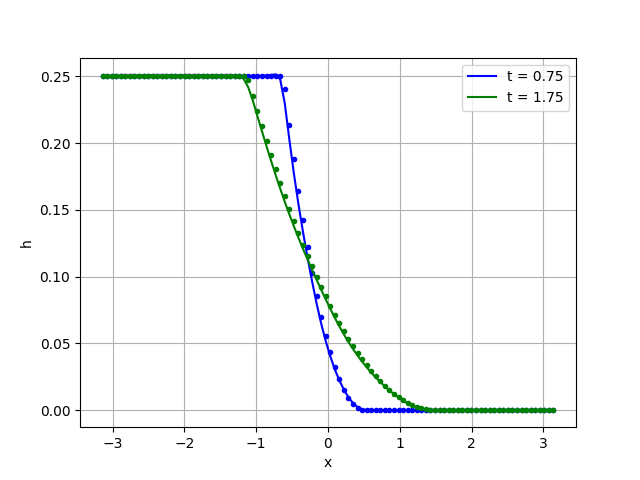}
\end{subfigure}
    \caption{Numerical solution for the dam break problem using physics-informed DeepONets (solid line) compared to the exact solution (dots). \textit{Left:} Solution for values $h_0=0.75$, $x_0=0.25$, $t_0=0.25$. \textit{Right:} Solution for values $h_0=0.25$, $x_0=-0.25$, $t_0=0.75$.}
    \label{fig:DamBreak}
\end{figure}

Figure~\ref{fig:DamBreak} shows the numerical results for the trained DeepONet, using two unseen initial conditions drawn from the same parameter space for $h_0$ and $x_0$, with an initial condition given by the exact solution for two particular values of $t_0$. These results verify the success of the DeepONet to learning the solution operator for this particular parameter space. That is, for obtaining these solutions, we only need to carry out an inference step on the trained neural network, thus alleviating the need for re-training the network for changing initial conditions.

Once a physics-informed DeepONet is trained, not only can it be applied to a new initial condition from the parameter space of the problem for which it had been trained, it can also be used iteratively for time-stepping~\cite{wang23a}. That is, the solution at the the final time $t=\Delta t$ can be used as an initial condition for the DeepONet again, making longer time integrations feasible that typically fail using physics-informed neural networks. This renders physics-informed DeepONets similar to standard numerical methods, albeit being able to use much larger time-steps than would be feasible with classical discretization methods.

\subsubsection{Swirling flow in a parabolic basin}\label{subsubSec:SwirlingBowl2D}

This two-dimensional benchmark follows another exact solution presented in~\cite{thac81a}, namely for the case of a rotating fluid with a flat water surface. Specifically, the bottom topography is again a parabolic function of the form $b=b_0(r^2/L^2-1)$, where $r=\sqrt{x^2+y^2}$, and the exact solution being
\begin{align*}
  &h_{\rm a}(t,x,y) = 2\eta\frac{b_0}{L}\left(\frac{x}{L}\cos\omega t-\frac{y}{L}\sin\omega t - \frac{\eta}{2L}\right)\\
  &u_{\rm a}(t,x,y) = -\eta \omega \sin\omega t,\\
  &v_{\rm a}(t,x,y) = -\eta\omega\cos\omega t,
\end{align*}
where $\omega=\sqrt{2gb_0}/L$, and $b_0$, $L$ and $\eta$ are the parameters of the problem. In the following, we use $b_0=1$, and $L=1$, and train a physics-informed DeepONet for the class of problems with varying values for $\eta$. 

Specifically, we sample $\eta$ from a random uniform distribution over the interval $[0.25,0.5]$, and using the exact solution as initial condition for various values of $t$, we then train a physics-informed operator network to learn the solution operator over the temporal interval $[0,\Delta t]$, with $\Delta t=2\pi/(4\omega)$. A total of 50,000 initial conditions from this class of problems is drawn.

\begin{figure}[!ht]
\centering
\begin{subfigure}[b]{\textwidth}
\includegraphics[width=0.33\textwidth]{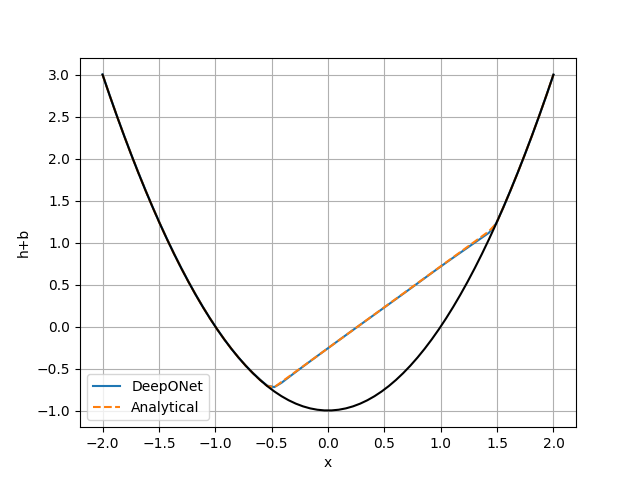}
\includegraphics[width=0.33\textwidth]{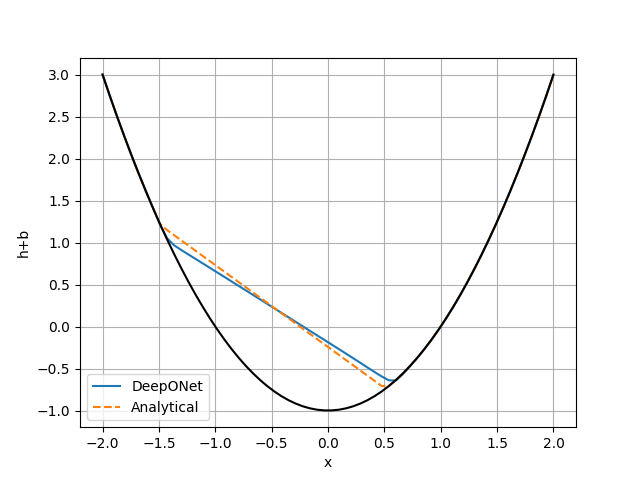}
\includegraphics[width=0.33\textwidth]{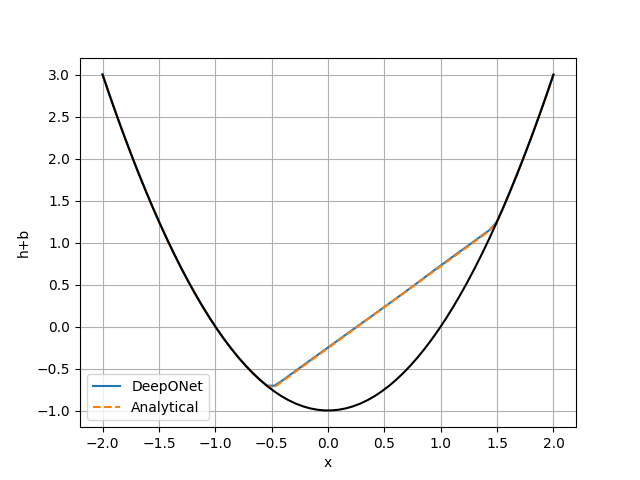}
\end{subfigure}
\caption{Oscillatory flow in a two-dimensional parabolic basis using a physics-informed DeepONet. The period of the solution is $2\pi/\omega$ and the results are shown at $t=0$ \textit{(left)}, $t=\pi/\omega$ \textit{(middle)} and $t=2\pi/\omega$ \textit{(right)} at the cross-section $y=0$.}
\label{fig:SwirlingBowl2D}
\end{figure}

Figure~\ref{fig:SwirlingBowl2D} shows the results of the trained physics-informed DeepONet, using an unseen new initial condition from the parameter range for which the DeepONet has been trained. Here we also illustrate the time-stepping capabilities of the trained DeepONet for 5 time steps, which cover a full period of the oscillating solution. Note that classical numerical methods would require far more than 5 time steps to capture a full period of the swirling bowl benchmark case in a numerically stable manner.

\subsection{Discussion}

Sections~\ref{subSec:PINNs} and~\ref{subSec:DeepONets} verified for several key benchmarks for tsunami inundation modeling that both physics-informed neural networks and DeepONets can recover high-fidelity solutions to the shallow-water equations with bottom topography. While physics-informed neural networks are easier to train compared to DeepONets, as they are optimized to satisfy one particular initial--boundary value problem rather than a class of such problem to learn the solution operator itself, they require re-training for each change in the initial and/or boundary condition. Given that training neural networks is much more costly than evaluating them, for time-critical applications such as real-time tsunami inundation modeling, physics-informed DeepONets would be more suitable compared to standard physics-informed neural networks. 

To give a rough computational comparison against a traditional numerical approach we compare the computational times require for training and inference for physics-informed neural networks and DeepONets against the meshless RBF-FD method developed in~\cite{brec18a}. The latter requires the generation of finite difference derivative operators, which will have to be re-generated each time the computational mesh changes, but can otherwise be computed offline and then re-used for varying initial conditions. Table~\ref{tab:ComputationalTimes} compares the approximate computational times required to obtain solutions for the two-dimensional parabolic bowl test cases considered in this paper.

\begin{table}[!ht]
\renewcommand{\arraystretch}{1.1}
\begin{tabular}{c|cc}
     & NN training / \textcolor{lightgray}{Operator generation} & Inference / \textcolor{lightgray}{Time stepping}  \\
     \hline
     \hline
     \textcolor{lightgray}{RBF-FD method}~\cite{brec18a} & $\sim$ 15 min & $\sim$ 5 min \\
     \hline
     PINNs & $\sim$ 1 h & $\sim$ 1 sec\\
     \hline
     DeepONets & $\sim$ 3 h & $\sim$ 10 sec\\
     \hline
\end{tabular}
\caption{Approximate computational times for obtaining the numerical solution to the two-dimensional parabolic bowl test benchmarks considered here. \textit{Operator generation} refers to the computational time required to compute the finite difference operators for the meshless RBF-FD method~\cite{brec18a}, while \textit{time stepping} refers to the computation time it takes to compute the solution up to~$t_{\rm f}$.}
\label{tab:ComputationalTimes}
\end{table}

We should like to point out here that a fair and quantitative comparison between physics-informed neural networks and traditional numerical methods is challenging for several reasons. Physics-informed neural networks require the use of GPUs, while traditional numerical methods typically are designed for CPUs. The number of collocation points is typically not comparable to the number of grid points used in classical discretization schemes~\cite{bihl22a}, and the notation of convergence in machine learning based methods is far less understood compared to traditional discretization schemes, as the former typically require stochastic optimization methods for multi-task learning which is known to be challenging~\cite{sene18a}; as such, the cost per accuracy ratio is much more difficult to assess for machine learning based techniques than for traditional discretization methods from scientific computing.

With that being said, it is typically the case that the inference cost of trained neural networks is only a fraction of the computational cost required for time-stepping in traditional numerical methods, see e.g.~\cite{muli22a} for a discussion in this regard for tsunami inundation modeling. We find this is the case as well for our trained neural networks, both standard physics-informed neural networks and DeepONets. As indicated in Section~\ref{sec:PINNs}, a main advantage of physics-informed DeepONets over physics-informed neural networks is that the former do not require re-training for changing initial conditions. We should also like to note that training physics-informed deep operator networks is possible for $\Delta t$ much larger than time steps that would be numerically stable in standard numerical schemes. For example, for the dam break problem considered in Section~\ref{subsubSec:DamBreakProblem} we used a time step of $\Delta t=1$, which would be infeasible for classical numerical methods, as would be the time step of $\Delta=2\pi/(4\omega)$ for the swirling bowl benchmark of Section~\ref{subsubSec:SwirlingBowl2D}. As such, while each step evaluating the deep operator network is potentially more expensive than a single step with a standard numerical method, especially for more complicated neural network architectures with several thousand parameters, much fewer of such steps are required to cover the same computational domain. This makes deep operator networks potentially attractive for time-critical applications such as real-time tsunami modeling.

\section{Conclusions}\label{sec:Conclusions}

We have introduced the use of physics-informed neural networks for tsunami modeling. The method is entirely meshless and does not require the use of a specific inundation model, making it thus ideal for capturing the wetting and drying process of wave--land interactions. We have illustrated with several classical benchmarks that physics-informed neural networks are capable of accurately modeling shallow-water wave inundation. We have illustrated the method here for the shallow-water equations but it could be readily adopted for other models used for tsunami inundation modeling~\cite{marr20a}.

While standard physics-informed neural networks learn a global solution interpolant for a particular initial--boundary value problem, and thus have to be re-trained once the initial or boundary conditions change, we have also shown that it is possible to learn the solution operator for the shallow-water equations with variable bottom topography directly. This has the advantage that training, while potentially costly, only has to be done once offline, and the trained model can be then be evaluated for new initial conditions in a straightforward and computationally inexpensive inference step. This not only speeds up obtaining new particular solution, but also allows for time-stepping to obtain solutions for longer time intervals than are otherwise feasible for physics-informed neural networks. 

As rapid simulation results are required for real-time tsunami hazard prediction, the physics-informed operator learning approach discussed here could potentially provide a viable alternative to standard numerical methods in use today. Specifically, as a next step towards assessing the real-world applicability of physics-informed deep operator networks for tsunami inundation modeling, one could attempt to learn the solution operators for an ensemble of numerical shallow-water solutions driven by simulations from realistic megathrust events \cite{haye18a,mai02a}. In comparison to other machine learning based approaches discussed in~\cite{liu21a}, which relied on learning the runup in isolated points for which a purely driven-driven neural network has been trained, the trained physics-informed DeepONets would allow evaluation of the near-field and runup of tsunamis over the entire spatio-temporal computational domain of interest for one particular coastline. We hope this would allow combining the merits of purely data-driven machine learning approaches, which have been pursued in the literature in recent years, with those from traditional numerical approaches solving the governing equations of shallow-water waves. We aim to carry out this task in a subsequent investigation.

\begin{ack}
This research was undertaken, in part, thanks to funding from the Canada Research Chairs program, the NSERC Discovery Grant program, and the Deutsche Forschungsgemeinschaft (DFG, German Research Foundation) -- Project-ID 274762653 -- TRR 181.
\end{ack}

{\footnotesize\setlength{\itemsep}{0ex}

\end{document}